\documentclass[aps,superscriptaddress,amsmath,amssymb,twocolumn,showpacs,floatfix,english,noshowpacs]{revtex4}
\usepackage{url}
\usepackage{bm,bbm}
\usepackage{graphicx,lipsum}
\usepackage[dvipsnames]{xcolor}
\usepackage[colorlinks=true, urlcolor=blue, linkcolor=blue, citecolor=blue, pdftex]{hyperref}
\usepackage[english]{babel}

\usepackage{soul}
\usepackage{blindtext}
\usepackage{amsmath}
\usepackage{amsfonts}
\usepackage{mathrsfs}
\usepackage{bbold}
\usepackage[footnote=true]{snotez}
\usepackage{braket}

\newcommand{\dagga}{{\phantom{\dagger}}}

\begin{document}

\title{A real-space many-body marker for correlated ${\mathbb Z}_2$ topological insulators}

\author{Ivan Gilardoni}
\affiliation{Dipartimento di Scienza e Alta Tecnologia, Universit\`a dell'Insubria, Via Valleggio 11, I-22100 Como, Italy}
\author{Federico Becca}
\affiliation{Dipartimento di Fisica, Universit\`a di Trieste, Strada Costiera 11, I-34151 Trieste, Italy}
\author{Antimo Marrazzo}
\affiliation{Dipartimento di Fisica, Universit\`a di Trieste, Strada Costiera 11, I-34151 Trieste, Italy}
\author{Alberto Parola}
\affiliation{Dipartimento di Scienza e Alta Tecnologia, Universit\`a dell'Insubria, Via Valleggio 11, I-22100 Como, Italy}

\date{\today}

\begin{abstract}
Taking the clue from the modern theory of polarization [R. Resta, Rev. Mod. Phys. {\bf 66}, 899 (1994)], we identify an operator to distinguish 
between ${\mathbb Z}_2$-even (trivial) and ${\mathbb Z}_2$-odd (topological) insulators in two spatial dimensions. Its definition extends the 
position operator [R. Resta and S. Sorella, Phys. Rev. Lett. {\bf 82}, 370 (1999)], which was introduced in one-dimensional systems. We first 
show a few examples of non-interacting models, where single-particle wave functions are defined and allow for a direct comparison with standard 
techniques on large system sizes. Then, we illustrate its applicability for an interacting model on a small cluster, where exact diagonalizations 
are available. Its formulation in the Fock space allows a direct computation of expectation values over the ground-state wave function (or any 
approximation of it), thus allowing us to investigate generic interacting systems, such as strongly-correlated topological insulators.
\end{abstract}

\maketitle

{\it Introduction.} Topological insulators represent nowadays a pillar of condensed-matter physics~\cite{hasan2010,qi2011}, defining a class of 
materials that are fundamentally distinct from ordinary band insulators. Their history originates in the early days of the integer quantum Hall 
effect, where topology plays a prominent role~\cite{thouless1982,haldane1988}. Here, time-reversal symmetry is broken and different quantum 
states are possible, which can be distinguished by the total Chern number of occupied bands. This leads to a ${\mathbb Z}$ classification of 
distinct topological states in two dimensions, different from conventional band insulators. A major step forward has been achieved when it was 
realized that enforcing the time-reversal symmetry the situation changes radically~\cite{kane2005}. In this case, only two possibilities are 
left, thus leading a ${\mathbb Z}_2$ classification, where trivial (${\mathbb Z}_2$-even) and topological (${\mathbb Z}_2$-odd) states exist.
Their full characterization has been obtained in non-interacting systems, where the inspection of Bloch or Wannier wave functions allows a 
straightforward determination of their properties~\cite{vanderbilt}. For example, trivial and topological states can be distinguished by looking 
at the time-reversal polarization, which can be computed in terms of Wannier centers~\cite{fu2006,soluyanov2011,gresch2016}. In addition, 
whenever inversion symmetry is present, the computation is reduced to the determination of the parity of occupied states at time-reversal 
momenta~\cite{fu2007}.

The inclusion of electron-electron interaction, beyond simple mean-field approximations, is far from being simple and straighforward. Indeed, the
analysis based upon single-particle wave functions is lost, forcing us to deal with the many-body state in its entirety. Since the early studies 
on the integer quantum Hall effect, Niu and collaborators proposed an ingenious way to compute topological observables (e.g., Chern numbers) by 
averaging over boundary conditions suitable derivatives of the many-body wave function~\cite{niu1985}. As originally noted in Ref.~\cite{niu1985} 
and recently verified numerically~\cite{kudo2019}, it turns out that the integration is actually not necessary and the Berry curvature evaluated
by computing the derivatives at fixed boundary conditions is already quantized. However, this procedure is not easily implemented, since it 
requires the determination of the ground state for different choices of the boundary conditions~\cite{sheng2003,wan2005,hafezi2008}, which is
particularly difficult when dealing with approximate solutions of the model.

Recently, a few investigations focused on the Bernevig-Hughes-Zhang (BHZ) model~\cite{bernevig2006} on the lattice, supplemented with Hubbard-like 
interactions, to determine their effects on the transition between trivial and topological insulators~\cite{amaricci2015,amaricci2016,barbarino2019}. 
In one spatial dimension, density-matrix renormalization group (DMRG) can be used~\cite{barbarino2019} to evaluate the local spin at the system 
edges, whose presence provides an indication on the topological nature of the ground state. However, this procedure is not fully satisfactory, in 
the view of defining a marker that can unambiguously distinguish the two band insulators. Alternatively, some approximate method can be used, as 
for example dynamical mean-field theory, to investigate either two- or three-dimensional systems~\cite{amaricci2015,amaricci2016}. Here, trivial and 
topological insulators are discriminated on the basis of the low-energy behavior of the electron self-energy~\cite{wang2012}, which is not easily 
accessible within other ground-state approaches (e.g., DMRG or quantum Monte Carlo methods). Real-space Chern markers have been also introduced 
for non-interacting systems~\cite{bianco2011} and extended, within dynamical mean-field theory, to include the effects of electron-electron 
interactions~\cite{amaricci2017}. 

Well before these developments in the framework of topological insulators, Resta and Sorella~\cite{resta1999} introduced a many-body operator to 
discriminate metals and insulators in interacting systems. Building on the modern theory of polarization~\cite{resta1994}, they focused the 
attention on one-dimensional models, defining
\begin{equation}\label{eq:z1}
 {\hat Z} = \exp \left ( \frac{2\pi i}{M} \sum_{j=1}^{M} x_j \, {\hat n}_j \right ),
\end{equation}
where $M$ is the number of sites and ${\hat n}_j$ is the electron density operator on the $j^{th}$ site, whose physical coordinate is $x_j$. Then, 
the {\it modulus} of its expectation value over the (normalized) ground state $z=\langle \Psi_0|{\hat Z}|\Psi_0 \rangle$ can be used to measure the 
localization length $\lambda^2=-[M/(2\pi)]^2 \ln |z|^2$. In the thermodynamic limit, a metal is characterized by $\lambda \to \infty$ ($|z| \to 0$) 
and an insulator by a finite $\lambda$ ($|z| \to 1$). For insulators, the phase of $z$ is related to electronic polarization (in units of the 
electric charge $e$) through the many-body Berry phase $\gamma$ \cite{resta1998,resta2018}:
\begin{equation}
P =  \frac{\gamma}{2\pi} = \frac{1}{2\pi} {\rm  Im} \ln z.
\end{equation}

The ${\hat Z}$ operator is very useful to detect the Mott transition in the one-dimensional Hubbard model~\cite{capello2005,motta2020}. However, 
quite remarkably, little attention has been given to the phase (or sign, for centrosymmetric lattices) of $z$. In fact, two classes of interacting 
centrosymmetric insulators may be distinguished by having either $z=1$ or $-1$~\cite{restanotes}. While in the one-dimensional non-interacting case
the topological properties of the Berry phase have been already discussed in relation to the surface charge theorem~\cite{king1993,kudin2007}, the 
phase of $z$ in interacting systems has not been investigated. In addition, since then, no attempts to extend the analysis to two-dimensional systems 
have been pursued.

In this Letter, we perform an important step forward in this direction, defining a marker, which is inspired by Eq.~\eqref{eq:z1}; then, its phase
can be expressed in terms of the Chern number, allowing us to discriminate between ${\mathbb Z}_2$-even and ${\mathbb Z}_2$-odd insulators. Specific 
examples of non-interacting lattice systems with time-reversal symmetry, like the BHZ~\cite{bernevig2006} or the Kane-Mele (KM)~\cite{kane2005} models 
are provided. In addition, calculations on the interacting BHZ model (where the on-site Hubbard-$U$ is included) are also reported for a $3 \times 3$ 
cluster (with $18$ electrons). The present work will allow one to perform ground-state calculations in interacting systems and obtain a clearcut 
way to distinguish trivial and topological states.

{\it Settings and definitions.} In the following, we will focus on two-band lattice models of spinful fermions at half filling (i.e., with two 
electrons per site on average). In the BHZ model, there are two orbitals, labelled by $\eta=\pm$, on each site ${\bf R}$ of the underlying Bravais 
lattice; instead, in the KM model, there are two sites, again labelled by $\eta=\pm$, in the unit cell. The band structure is assumed to display a 
gap, leading to an insulating ground state. We will first consider Hamiltonians conserving the $z$ projection of the total spin ${\hat S}_z$, which 
is customary in the literature~\cite{rachel2018}. In this case, the $\mathbb{Z}_2$ invariant can be equivalently discussed in terms of the parity 
of the spin Chern number~\cite{prodan2009}, calculated over the occupied states with spin up (or down) only~\cite{vanderbilt}. The effects of 
symmetry-breaking perturbations are also discussed. 

In order to generalize the definition of lattice position operator of Eq.~\eqref{eq:z1} in finite clusters of any geometry and dimension, we have 
to introduce a many-body operator that commutes with lattice translations and contains the information on the average electron position. A useful 
definition, which plays a central role in our treatment, is given by
\begin{equation}\label{eq:hatz}
{\hat Z}_{\sigma}(\delta {\bf k}) = \exp \left (i \delta {\bf k} \cdot \sum_j {\bf R}_j\,{\hat n}_{j,\sigma} \right ),
\end{equation}
where $\delta {\bf k}$ is a yet unspecified wavevector, quantized according to the lattice geometry, and 
${\hat n}_{j,\sigma}=\sum_{\eta} {\hat n}_{j,\eta,\sigma}$ is the spin-projected electron density operator on the $j^{th}$ Bravais lattice site, 
located at ${\bf R}_j$. Whatever choice of the parameter $\delta {\bf k}$, the operator ${\hat Z}_{\sigma}(\delta {\bf k})$ is a legitimate estimator 
of the average electron position within the cluster.

\begin{figure}
\includegraphics[width=0.96\linewidth]{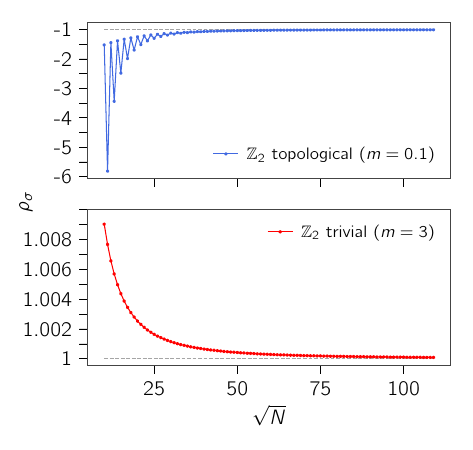}
\caption{\label{fig1}
The quantity $\rho_{\sigma}$ for the BHZ model with $t=\lambda=1$ as a function of the linear lattice size $L=\sqrt{N}$. In the upper panel $m=0.1$, 
corresponding to a $\mathbb{Z}_2$ topological insulator; in the lower panel $m=3$, corresponding to a trivial insulator. Notice that in the BHZ
model $\rho_{\sigma}$ is real.}
\end{figure}

It is useful to first prove an interesting property of the operator defined in Eq.~\eqref{eq:hatz}. Indeed, the ground-state average of 
${\hat Z}_{\sigma}(\delta {\bf k})$ can be related to the overlap between the ground states of the model with and without a magnetic field piercing 
the two-dimensional torus. This can be easily proved by noting that the unitary operator ${\hat Z}_{\sigma}(\delta {\bf k})$ implements a gauge 
transformation on the fermion operators: 
\begin{equation}
{\hat Z}^\dagga_{\sigma}(\delta {\bf k}) \,{\hat c}^\dagger_{j,\eta,\sigma} \,{\hat Z}^\dagger_{\sigma}(\delta {\bf k}) = 
e^{i\delta {\bf k} \cdot {\bf R}_j} \, {\hat c}^\dagger_{j,\eta,\sigma},
\end{equation}
where ${\hat c}^\dagger_{j,\eta,\sigma}$ creates an electron on the Bravais site $j$, orbital $\eta$, and spin $\sigma$. Then, if $|\Psi_0\rangle$ 
is the many-body ground state of the Hamiltonian ${\hat {\cal H}}$ with periodic boundary conditions, then 
$|\Psi_0(\delta {\bf k})\rangle = {\hat Z}_{\sigma}(\delta {\bf k})\,|\Psi_0\rangle$ is the ground state of the Hamiltonian 
${\hat Z}_{\sigma}(\delta {\bf k}) \, {\hat {\cal H}} \, {\hat Z}^\dagger_{\sigma}(\delta {\bf k})$. The density operators in the transformed 
Hamiltonian are left invariant by the gauge transformation, while the hopping terms of the electrons with spin $\sigma$ acquire a phase factor 
which can be attributed, via the Peierls substitution, to the presence of a (pure) gauge field, i.e., the presence of an integer number of magnetic 
quantum fluxes piercing the torus. Notice that periodic-boundary conditions are preserved by the gauge transformation due to the quantization of 
$\delta {\bf k}$. As a result, the ground-state average of ${\hat Z}_\sigma(\delta {\bf k})$ {\it without} the quantum flux, equals the overlap 
between the ground states of the model {\it with} and {\it without} the quantum flux:
\begin{equation}\label{eq:gauge}
\langle \Psi_0| {\hat Z}_{\sigma}(\delta {\bf k})|\Psi_0\rangle = \langle \Psi_0|\Psi_0(\delta {\bf k})\rangle.
\end{equation}
This relation is exact but rests upon the precise definition of the phase factor of the ground state $|\Psi_0(\delta {\bf k})\rangle$, which must 
be chosen according to the previous derivation. 

The analysis of non-interacting centrosymmetric models in one spatial dimension provides a useful check on the method, as we show in the Supplemental 
Material~\cite{suppmat}. 

{\it The BHZ model.} Let us focus now on the BHZ model, defined on a a square lattice with $N=L \times L$ sites by the Hamiltonian 
${\hat {\cal H}} = \sum_{\sigma} {\hat {\cal H}}_{\sigma}$ with
\begin{eqnarray} 
 {\hat {\cal H}}_{\sigma} &=& -\frac{t}{2} \sum_{\langle i,j\rangle,\eta} \eta {\hat c}^\dagger_{i,\eta,\sigma} {\hat c}^\dagga_{j,\eta,\sigma}
 +m \sum_{i,\eta} \eta {\hat c}^\dagger_{i,\eta,\sigma} {\hat c}^\dagga_{i,\eta,\sigma} \nonumber \\
 &-&\frac{\lambda}{2} \sum_{\langle i,j\rangle} e^{i \varphi^{\sigma}_{ij}} {\hat c}^\dagger_{i,+,\sigma} {\hat c}^\dagga_{j,-,\sigma} + {\rm H.c.},
\label{eq:BHZ}
\end{eqnarray}
where $\langle i,j\rangle$ are nearest-neighbor sites and the phase factor $\varphi^{\uparrow}_{ij}=-\varphi^{\downarrow}_{ij}$ depends on the 
vector ${\bf r}_{ij}={\bf R}_j-{\bf R}_i$, i.e., $\pm \pi/2$ for ${\bf r}_{ij}=(\pm 1,0)$, $0$ for ${\bf r}_{ij}=(0,1)$, and $\pi$ for 
${\bf r}_{ij}=(0,-1)$.

Here, we take $\delta k_x = \delta k_y = \frac{2\pi}{L}$ and introduce the ratio 
\begin{equation}\label{eq:rhobhz}
\rho_{\sigma} = \frac{\langle \Psi_0| {\hat Z}_{\sigma}(\delta k_x,\delta k_y)|\Psi_0\rangle}
{\langle \Psi_0|  {\hat Z}_{\sigma}(\delta k_x,0)|\Psi_0\rangle \, \langle \Psi_0| {\hat Z}_{\sigma}(0,\delta k_y)|\Psi_0\rangle}.
\end{equation}
Then, we can exploit Eq.~\eqref{eq:gauge} in order to express $\rho_{\sigma}$ in terms of overlaps, independently of the chosen global phase of 
the ground states:
\begin{equation}
\rho_{\sigma} = \frac{\langle \Psi_0(-\delta k_x,0)|\Psi_0(0,\delta k_y)\rangle}
{\langle \Psi_0(-\delta k_x,0)|\Psi_0\rangle \, \langle\Psi_0|\Psi_0(0,\delta k_y)\rangle}.
\end{equation}
For a non-interacting model, the ground state is written as a Slater determinant of the single-particle eigenfunctions and each overlap is written 
as the determinant of the matrix built out of the overlaps of the single-particle states. Expressing each single-particle eigenstate in the Bloch 
form, the determinant can be explicitly evaluated in the thermodynamic limit as: 
\begin{widetext}
\begin{eqnarray}
 && \rho_{\sigma} = \prod_q \, \frac{\langle u_{q+\delta k_x,\sigma}|u_{q-\delta k_y,\sigma}\rangle}
 {\langle u_{q+\delta k_x,\sigma}|u_{q,\sigma}\rangle \, \langle u_q|u_{q-\delta k_y,\sigma}\rangle} =
 \prod_q \, \left [ 1 - \delta k_x\delta k_y \left ( \langle \partial_{q_x} u_{q,\sigma}|\partial_{q_y} u_{q,\sigma}\rangle -  
 \langle \partial_{q_x} u_{q,\sigma} |u_{q,\sigma}\rangle \, \langle u_{q,\sigma}|\partial_{q_y} u_{q,\sigma}\rangle \right ) \right ] \nonumber \\
 &&= \exp \left [ -\int_{BZ} dq_x dq_y \,  \left ( \langle \partial_{q_x} u_{q,\sigma}|\partial_{q_y} u_{q,\sigma}\rangle -  
 \langle \partial_{q_x} u_{q,\sigma}|u_{q,\sigma}\rangle \, \langle u_{q,\sigma}|\partial_{q_y} u_{q,\sigma}\rangle \right ) \right ]
 = |\rho_{\sigma}| \exp \left (i\pi C_{\sigma} \right ).
\end{eqnarray}
\end{widetext}
Therefore, $\rho_{\sigma}$ is expressed in terms of the off-diagonal components of the metric-curvature tensor~\cite{resta2018} and its phase 
is written as the integral of the Berry curvature, implying that it is just $\pi$ times the spin Chern number of the occupied spin-$\sigma$ 
manifold. The results of $\rho_{\sigma}$ as a function of the lattice size $L$ are shown in Fig.~\ref{fig1}, for two cases, corresponding to 
trivial and topological insulators. For this model, $\rho_{\sigma}$ is real on finite clusters and its modulus tends to $1$ in the thermodynamic 
limit. We emphasize that the sign of $\rho_{\sigma}$ provides a clear marker for the topological nature of the ground state, since it does not
depend on the cluster size.

\begin{figure*}
\includegraphics[width=0.44\linewidth]{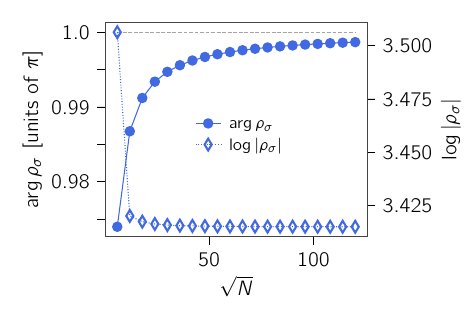}
\includegraphics[width=0.44\linewidth]{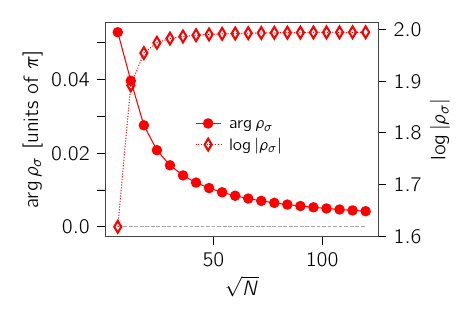}
\caption{\label{fig2}
Modulus (empty points) and phase (full points) of $\rho_{\sigma}=|\rho_{\sigma}|\,e^{i\phi}$ in the KM model of Eq.~\eqref{eq:hkm} in absence 
of Rashba coupling ($V_R=0$), as a function of the number of sites $N$ for $t=1$ and $V_{SO}=1/3$. In the left panel $m=1$ and the system is a 
$\mathbb{Z}_2$ topological insulators, while in the right panel $m=2$ and the system is a trivial insulator.}
\end{figure*}

\begin{figure*}
\includegraphics[width=0.44\linewidth]{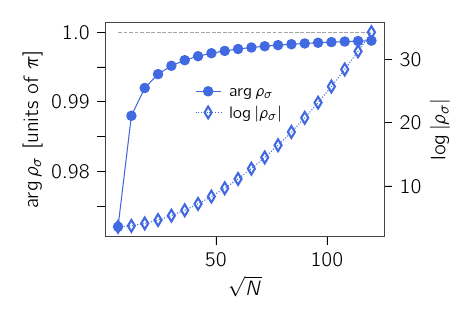}
\includegraphics[width=0.44\linewidth]{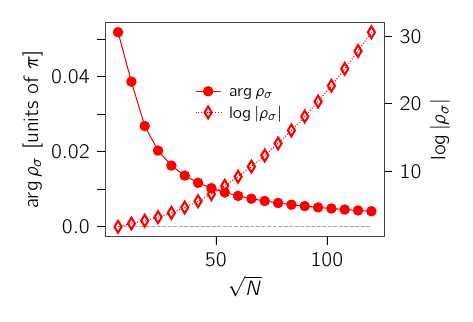}
\caption{\label{fig3}
The same as in Fig.~\ref{fig2} with a finite Rashba coupling $V_R=0.1$.}
\end{figure*}

{\it The KM model.} We now consider the Kane-Mele model, defined in the honeycomb lattice with $2N$ sites, which are labelled by 
$\alpha=(i,\eta)$ (where $i$ denotes the unit cell of the Bravais lattice and $\eta$ the site in the unit cell). The Hamiltonian is given by:
\begin{eqnarray}
 {\hat {\cal H}} &=& -t \sum_{\langle \alpha,\beta\rangle} {\hat c}^\dagger_{\alpha} {\hat c}^\dagga_{\beta} 
 + \frac{2i}{\sqrt{3}} V_{SO} \sum_{\langle\langle \alpha,\beta\rangle\rangle} 
 {\hat c}^\dagger_{\alpha} \,{\bf \sigma} \cdot ({\bf d}_{\gamma,\beta} \times {\bf d}_{\alpha,\gamma}) {\hat c}^\dagga_{\beta} \nonumber \\
 &+& \sum_{\alpha} m_{\alpha} {\hat c}^\dagger_{\alpha} {\hat c}^\dagga_{\alpha} 
 +i V_R \sum_{\langle \alpha,\beta \rangle} {\hat c}^\dagger_{\alpha}\, \hat {\bf z} \cdot ({\bf \sigma} \times {\bf d}_{\alpha,\beta}) {\hat c}^\dagga_{\beta} 
\label{eq:hkm}
\end{eqnarray}
where $\langle \alpha,\beta\rangle$ and $\langle\langle \alpha,\beta\rangle\rangle$ are nearest and next-nearest neighbors in the honeycomb lattice, 
${\hat c}^\dag_{\alpha} = ({\hat c}^\dag_{\alpha,\uparrow},{\hat c}^\dag_{\alpha,\downarrow})$, ${\bf \sigma}=(\sigma_x,\sigma_y,\sigma_z)$ 
are Pauli matrices and ${\bf d}_{\alpha,\beta}$ is a vector pointing from site $\beta$ to $\alpha$, while $\gamma$ denotes the common nearest 
neighbor of the two next-nearest-neighbor sites $\alpha$ and $\beta$. The on-site (mass) term $m_{\alpha} =\pm m$ has alternate signs on each 
sublattice of the honeycomb lattice. 

In the absence of Rashba coupling ($V_R=0$), the total spin projection is still conserved by the Hamiltonian and the previous analysis is readily 
applicable. The quantized wavevectors $\delta {\bf k}$ appearing in the definition~\eqref{eq:hatz} must be chosen according to the quantization 
rules of the underlying triangular Bravais lattice. We choose $\delta {\bf k}_i$ as the smallest wavevector in the direction of the $i^{th}$ 
primitive vector of the reciprocal lattice; in a $L\times L$ cluster with primitive vectors ${\bf a}_1$ and ${\bf a}_2$, we have that 
$\delta {\bf k}_i \cdot {\bf a}_j = \frac{2\pi}{L}\,\delta_{ij}$. As done before, we now define 
\begin{equation}\label{eq:rhokm}
\rho_{\sigma} = \frac{\langle \Psi_0| {\hat Z}_{\sigma}(\delta {\bf k}_1+\delta {\bf k}_2)|\Psi_0\rangle}
{\langle \Psi_0| {\hat Z}_{\sigma}(\delta {\bf k}_1)|\Psi_0\rangle \, \langle \Psi_0| {\hat Z}_{\sigma}(\delta {\bf k}_2)|\Psi_0\rangle}.
\end{equation}
The derivation closely follows the one sketched before. Here, $\rho_{\sigma}$ is complex on any finite sizes but becomes real for $N \to \infty$, 
where its phase equals $\pi$ times the Chern number of the band. Therefore, it represents an easily computable quantity that marks the topological 
transition. The numerical results for the KM model in absence of Rashba coupling are reported in Fig.~\ref{fig2}. Even if our marker~\eqref{eq:rhokm} 
is complex on any finite cluster, its phase is very close to either $0$ or $\pi$, even on small sizes. Therefore, the identification of the topological 
nature of the ground state can be easily assessed.

When the Rashba coupling $V_R$ is included, the $z$ component of the total spin is no longer a conserved quantity. Still, we keep the same formal 
definition of $\rho_{\sigma}$ in Eq.~\eqref{eq:rhokm} and show that its phase remains quantized even for $V_R>0$. The numerical results are reported 
in Fig.~\ref{fig3}. For large $N$ the imaginary part gets smaller and eventually tends to zero while the modulus $|\rho_{\sigma}|$ diverges in the 
thermodynamic limit; however, its phase $\phi$ can be again taken as a marker for the ${\mathbb Z}_2$ topological transition. In fact, the convergence 
of the phase of $\rho_{\sigma}$ to $\pi$ ($0$) in the topological (trivial) phase is not affected by the presence of Rashba coupling.

\begin{figure}
\includegraphics[width=0.96\linewidth]{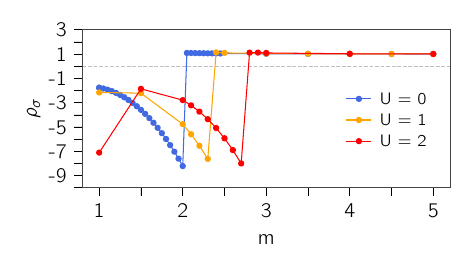}
\caption{\label{fig4}
The many-body $\mathbb{Z}_2$ marker $\rho_{\sigma}$ for the interacting BHZ model with $t=\lambda=1$ in the $3 \times 3$ cluster with $18$ 
electrons, as obtained by exact diagonalization. The results are shown for $U=1$ and $2$, as a fuction of the on-site term $m$. The case with 
$U=0$ (on the same cluster) is also reported for comparison. As in Fig.~\ref{fig1}, $\rho_{\sigma}$ is real.}
\end{figure}

{\it The interacting BHZ model.} Here, we add the Hubbard-$U$ interaction between electrons with the same orbital $\eta$ in the BHZ model of
Eq.~\eqref{eq:BHZ}, namely, ${\cal H}_{\rm int}=U \sum_{j,\eta} {\hat n}_{j,\eta,\uparrow} {\hat n}_{j,\eta,\downarrow}$~\cite{barbarino2019}. 
The ground state can no longer be written in terms of a single Slater determinant and many-body methods are necessary to evaluate expectation 
values, as that in Eq.~\eqref{eq:rhobhz}. Here, we perform exact diagonalizations on a $3 \times 3$ cluster with $18$ electrons to give a proof 
of concept for the applicability of the many-body marker that we introduced. In Fig.~\ref{fig4}, we report the results for $U=1$ and $2$ 
(with $\lambda=1$) by varying the on-site term $m$, the non-interacting case being also reported for comparison. The ground state is topological 
for small values of $m$ and the transition to the trivial insulator is marked by an abrupt jump from negative to positive values of 
$\rho_{\sigma}$. Notice that the presence of the Hubbard-$U$ interaction shifts the transition point from $m=2$ at $U=0$ to $m=2.75(5)$ at $U=2$, 
indicating that the electron-electron repulsion favors the topological phase.

{\it Conclusions.} In summary, we have examined the role of the spin-projected position operator ${\hat Z}_{\sigma}(\delta {\bf k})$ in the 
topological transition of lattice models, proving that it allows us to define a robust marker, whose phase clearly identifies the occurrence 
of a change in the topological properties of the ground-state wave function. The ${\hat Z}_{\sigma}(\delta {\bf k})$ operator is particularly 
suited for wave-function-based approaches (e.g., quantum Monte Carlo, Lanczos, and density-matrix renormalization group), where the topological
nature can be extracted even in small clusters. Other markers, have been introduced and employed in previous works~\cite{bianco2011,amaricci2017}. 
However, being based on a single-particle picture (e.g., by the use of Wannier orbitals), these markers can be exploited in dynamical mean-field 
theory investigations but their application to fully many-body states is not possible. Our ${\mathbb Z}_2$ marker bears some resemblance with the 
many-body invariant for Chern insulators discussed in Ref.~\cite{kang2021}; however, while the latter one needs calculations with different boundary 
conditions, our maker is defined by a {\it single} many-body computation. Most importantly, the definition of $\rho_{\sigma}$, can be exploited to 
study interaction-induced topological transitions in strongly-correlated electron models. The very same definition can be applied even if the total 
spin projection $S_z$ is not conserved, e.g., in presence of the Rashba coupling in the Hamiltonian. Finally, the position operator could be useful 
also in experimental setups on quantum gases trapped in optical lattices, in which high-resolution imaging is now possible~\cite{enders2011}, 
allowing a direct evaluation of operators like ${\hat Z}_{\sigma}(\delta {\bf k})$.

F.B. and A.P. would like to dedicate this work to the memory of Sandro Sorella, friend and colleague, whose seminal contribution inspired our 
investigation. We thank M.-F. Yang for having drawn our attention to Ref.~\cite{kang2021}.

\newpage

\section{Supplemental Material}
{\it One-dimensional models.} In a one-dimensional ring with $N$ elementary cells, we choose $\delta k = \frac{2\pi}{N}$. The ground-state
average of the operator~\eqref{eq:hatz} is easily evaluated in non-interacting models because the ground state $|\Psi_0\rangle$ can be expressed 
as a product of two Slater determinants of single-particle eigenfunctions of spin-$\sigma$ electrons: 
\begin{equation}
\psi_{q,\sigma}(R,\eta) = \frac{1}{\sqrt{N}}\,e^{iqR}\,u_{q,\sigma}(\eta), 
\end{equation}
where $(R,\eta)$ are respectively the site and the orbital of the electron. 
Moreover, ${\hat Z}_{\sigma}(\delta k)$ is a one-body operator that, acting on the Slater determinant 
of spin-$\sigma$ wave functions, gives another Slater determinant of the (new) single particle wave functions
\begin{equation}
	{\hat Z}_{\sigma}(\delta k)\,\psi_{q,\sigma}(R,\eta) = \frac{1}{\sqrt{N}}\,e^{i(q+\delta k)R}\,u_{q,\sigma}(\eta)
\end{equation}
Then, the overlap between the old and the new Slater determinants gives the determinant of the $N\times N$ overlap matrix
\begin{equation}
O_{p,q}=\langle\psi_{p,\sigma}| {\hat Z}_{\sigma}(\delta k)|\psi_{q,\sigma}\rangle=\delta_{p,q+\delta k} \langle u_{p,\sigma}|u_{q,\sigma}\rangle,
\end{equation}
which equals the product of its non-vanishing elements times $(-1)^{N-1}$. The overlap of the Bloch functions
\begin{equation}
\langle u_{q+\delta k,\sigma}|u_{q,\sigma}\rangle = \sum_{\eta=\pm} u_{q+\delta k,\sigma}^*(\eta)\,u_{q,\sigma}(\eta)
\end{equation}
can be explicitly evaluated for $N\gg 1$ as 
\begin{eqnarray}\label{eq:berry}
 && \langle \Psi_0| {\hat Z}_{\sigma}(\delta k)|\Psi_0 \rangle = (-1)^{N-1}
 \prod_q \, \left ( 1 + \delta k \,\langle \partial_q u_{q,\sigma}|u_{q,\sigma}\rangle \right ) \nonumber \\
         &&= (-1)^{N-1} \exp \left [ \int_{0}^{2\pi} dq \, \langle \partial_q u_{q,\sigma}|u_{q,\sigma}\rangle \right ].
\end{eqnarray}
Here, the overall sign that depends on the cluster size can be eliminated by changing the defintion of Eq.~\eqref{eq:hatz} by the 
substitution ${\hat n}_{j,\sigma} \to {\hat n}_{j,\sigma} - {\overline n}_{\sigma}$, where ${\overline n}_{\sigma}$ is the average electron density 
(per spin). The inner product is purely imaginary and then the average of the position operator is a phase factor. The latter one is given 
by the integral over the full Brillouin zone (i.e., from $0$ to $2\pi$) of the Berry phase of the occupied band. For centrosymmetric systems, 
the average of the spin projected position operator is either $+1$ or $-1$, according to the topological nature of the band structure. As 
such, the phase is quantized in integer multiples of $\pi$. The quantization of the phase of ${\hat Z}_{\sigma}(\delta k)$ is a robust feature 
with respect to the addition of interactions if particle-hole symmetry is preserved. 

\end{document}